# Partial antiferromagnetism in spin-chain $Sr_5Rh_4O_{12}$, $Ca_5Ir_3O_{12}$ and $Ca_4IrO_6$ single crystals

G. Cao, V. Durairaj, and S. Chikara,
Department of Physics and Astronomy
University of Kentucky, Lexington, KY 40506

S. Parkin
Department of Chemistry
University of Kentucky, Lexington, KY 40506

P. Schlottmann
Physics Department, Florida State University
Tallahassee, FL 32306

We report a structural, thermodynamic and transport study of the newly synthesized $Sr_5Rh_4O_{12}$, $Ca_5Ir_3O_{12}$ and $Ca_4IrO_6$ single crystals. These quasi-one-dimensional insulators consist of a triangular lattice of spin chains running along the c-axis, and are commonly characterized by a partial antiferromagnetic (AFM) order, a small entropy removal associated with the phase transitions and a sizable low-temperature specific heat linearly proportional to temperature. $Sr_5Rh_4O_{12}$ is defined by an AFM order below 23 K with strong evidence for an Ising character and two step-like transitions in isothermal magnetization leading to a ferrimagnetic state at 2.4 T and a ferromagnetic state at 4.8 T, respectively. $Ca_5Ir_3O_{12}$ and $Ca_4IrO_6$ are also antiferromagnetically ordered below 7.8 K and 12 K, respectively, and show an unusually large ratio of the Curie-Weiss temperature to the Neel temperature. In particular, $Ca_5Ir_3O_{12}$, which includes both $Ir^{4+}$ and $Ir^{5+}$ ions, reveals that only S=1/2 spins of the $Ir^{4+}$ ions are involved in the magnetic ordering whereas S=3/2 spins of the $Ir^{5+}$ ions remain disordered. All results suggest the presence of the geometrical frustration that causes incomplete long-range AFM order in these quasi-one-dimensional compounds.

## I. Introduction

Geometrical frustration occurs in materials consisting of triangle-related lattices where the competing energies are compatible, resulting in a degeneracy of ground states or an inability to minimize competing energies at low temperatures. It has been observed in two- and three-dimensional frustrating lattices such as kagome and pyrochlore systems. Quasi-one-dimensional structures combined with geometrical frustration are less common and give rise to complex excitations and novel magnetic order. Such behavior is manifested in Co based compounds such as $CsCoCl_3$ [1], $Ca_3Co_2O_6$ [2], $Ca_3CoRhO_6$ [3, 4] and $Ca_3CoIrO_6$ [5, 6]. Intriguing quantum phenomena displayed by these low dimensional materials have recently generated a great deal of interest and discussion [4-13 and refs. therein]. The central feature of these systems is the unusually strong correlation between lattice structure and spin coupling that dictates the magnetism. The spin chains always comprise alternating face-sharing $CoO_6$ octahedra and $CoO_6$ trigonal prisms running along the c-axis. The different crystalline electric fields (CEF) generate different spin states for Co ions leading to chains that have sites with alternating high and low spin states. These chains form a triangular lattice in the ab-plane that causes geometrical frustration and exotic magnetism. The Ising system $Ca_3Co_2O_6$, for instance, orders ferrimagnetically below 24 K with strong ferromagnetic (FM) intrachain coupling and weak antiferromagnetic (AFM) interchain coupling [2]. This feature together with the geometry of the triangular lattice brings about frustration and a partially ordered AFM phase with irreversible step-like magnetization [2-10]. In spite of intensive efforts, understanding these novel phenomena is still a profound challenge, and it is conspicuous that these phenomena have been found exclusively in 3d-based, i.e. Co-based, materials.

In cobaltites, the Co-ions can exist in more than one oxidation state, and for each valence a few spin-states are possible, e.g. low, intermediate and high spin-configurations. This is a consequence of the strong competition between the CEF and the Coulomb energies (Hund's rules). Magnetism is less common in 4d and 5d based materials than in 3d based compounds because of the more extended d-orbitals and the weaker Coulomb interactions. Among the 4d- and 5d-transition elements, Ru, Rh and Ir are candidates for such behavior. In ruthenates the expanded 4d-orbitals typically lead to a CEF that is large compared to the Hund's exchange, but in Rh- and Ir-based oxides these energies could be competitive. Recent studies have shown that layered iridates such as $BaIrO_3$ [14-19] and $Sr_{n+1}Ir_nO_{3n+1}$ with n=1 and 2 [20-25] share a number of unique features, for instance, high-temperature weak ferromagnetism and an insulating ground state with phase proximity of a metallic state due chiefly to strong electron-lattice coupling, which very often alters and distorts the metal-oxygen bonding lengths and angles, lifting degeneracies of $t_{2g}$ orbitals and thus precipitating possible orbital ordering [17,19, 24, 25]. In spite of these studies, no spin-chain systems with geometric frustration had ever been found in 4d and 5d based materials without a presence of an incomplete 3d-electron shell until now.

In this paper, we report results of structural, magnetic, specific heat and transport measurements of the newly synthesized single crystal $Sr_5Rh_4O_{12}$, $Ca_5Ir_3O_{12}$ and $Ca_4IrO_6$. All these compounds, which are insulating (below 300 K for $Ca_5Ir_3O_{12}$), share crystal structures that favor the formation of spin chains and a triangular lattice perpendicular to the spin chains, and feature a number of common characteristics, such as a partial AFM order, a small change in heat capacity or a small entropy removal associated with the phase transitions and a sizable low-temperature specific heat linearly proportional to temperature, etc. These materials also show distinct magnetic behavior that distinguishes each other. For $Sr_5Rh_4O_{12}$, the AFM transition occurs sharply at 23 K only along the c-axis and no magnetic anomaly is discerned in the ab-plane, suggesting an Ising characteristic. The magnetic results also suggest a strong FM intrachain coupling and a weak AFM interchain coupling. In addition, there are two step-like transitions in the c-axis isothermal magnetization that lead to a ferrimagnetic state with 1/3 of the saturation moment $M_s$ at a critical field $B^*=2.4$ T and a fully saturated FM state at $B_c=4.8$ T. On the other hand, both $Ca_5Ir_3O_{12}$ and $Ca_4IrO_6$ are also antiferromagnetically ordered below 7.8 K and 12 K, respectively, and exhibit an unusually large ratio of the Curie-Weiss temperature to the Neel temperature (>36 for $Ca_5Ir_3O_{12}$), a characteristic of two- and three-dimensional frustrating lattices. In particular, $Ca_5Ir_3O_{12}$, which includes both $Ir^{4+}$ and $Ir^{5+}$ ions, reveals that only S=1/2 spins of the $Ir^{4+}$ ions are involved in the magnetic ordering whereas S=3/2 spins of the $Ir^{5+}$ ions remain disordered. The electrical resistivity of $Ca_5Ir_3O_{12}$ reveals a crossover from a metallic state to an insulating at 300 K as temperature decreases and endorses the quasi-one-dimensional characteristic. Nevertheless, these 4d- and 5d-based spin-chain systems reveal a host of novel phenomena and offer a wide window into low-dimensional magnetism involving geometrically frustrated states.

**II. Experimental**

The single crystals studied were grown using flux techniques. They were grown in Pt crucibles from off-stoichiometric quantities of $IrO_2$ ($Rh_2O_3$), $CaCO_3$ ($SrCO_3$) and $CaCl_2$ ($SrCl_2$) mixtures with $CaCl_2$ ($SrCl_2$) being self flux. The mixtures were first heated above 1330 °C in a Pt crucible covered by a Pt cover, soaked for 25 hours, and slowly cooled at 2-3 °C/hour to 1200 °C and finally cooled to room temperature at 100°C/hour. The starting Ca(Sr):Ru ratio and the thermal treatments are critical and subtle for the formation of the crystals, particularly for $Ca_5Ir_3O_{12}$ and $Ca_4IrO_6$. Reflecting the quasi-one-dimensional characteristic, these crystals are either rod- or needle-shaped. The crystal structures of these crystals were determined from a small equidimensional fragment (0.05x0.05x0.05 $mm^3$) using Mo $K\alpha$ x-rays on a Nonius KappaCCD single crystal diffractometer. Heat capacity measurements were performed with a Quantum Design PPMS that utilizes a thermal-relaxation calorimeter operating in fields up to 9 T. Magnetic properties were measured using a Quantum Design SQUID magnetometer PMPS 7T LX. The high temperature resistivity was measured using a four-leads dc method and a Displex closed cycle cryostat (Advanced Research Systems DE202) capable for a continuous temperature ramping from 9 K to 900 K. All results proved to be highly reproducible.

## III. Results and Discussion

### A. Single crystal $Sr_5Rh_4O_{12}$

Refinements of the x-ray diffraction data reveal that the needle-shaped $Sr_5Rh_4O_{12}$ has an ordered but inversion twinned trigonal structure with a space group of P3c1 (158) and a mixed valence state of $Rh^{3+}$ and $Rh^{4+}$. An alternative description with space group P-3c1 (165) required a model that forced disorder of octahedral and trigonal prismatic oxygen atoms. The cell parameters are $a=b=9.6017(3)$ Å and $c=21.3105(8)$ Å. The central structural feature is the formation of chains that run along the c-axis and consist of face-sharing $RhO_6$ octahedra and $RhO_6$ trigonal prisms as shown in Fig.1. The $RhO_6$ trigonal prisms and $RhO_6$ octahedra alternate along the chains with a sequence of one trigonal prism and three octahedra. Of six chains in a unit cell, four chains are distorted similarly; the other two chains related to the Rh11 ion are distorted somewhat differently from the other four. The intrachain Rh-Rh bond for all six chains varies from ~2.5 Å to ~2.7 Å. For instance, the intrachain Rh-Rh bond for one of the six chains alternates from 2.747 Å (Rh4-Rh1), 2.744 Å (Rh1-Rh2), 2.578 Å (Rh2-Rh3) to 2.586 Å (Rh3-Rh4) (see Fig.1b). These uneven Rh-Rh bond distances correlate well with the different ionic sizes of $Rh^{3+}$ ($4d^6$) and $Rh^{4+}$($4d^5$), which are 0.665 Å and 0.600 Å, respectively. Accordingly, the sequence of the $Rh^{3+}$ and $Rh^{4+}$ ions in the chains is likely to be $Rh^{3+}$(o), $Rh^{3+}$(p), $Rh^{4+}$(o) and $Rh^{4+}$(o) where p stands for the $RhO_6$ trigonal prism, and o $RhO_6$ octahedra (see Fig.1b).

Since the Rh ions at different sites are subject to different CEF, these ions can have different spin states. The octahedral coordination favors a large crystal field splitting $\Delta_o$ between the three lower $t_{2g}$ and the two higher $e_g$ orbitals. Because $\Delta_o$ is normally larger than the Hund's rule energy, a low spin state is anticipated for the Rh ions in the $RhO_6$ octahedra, namely, S=0 and 1/2 for the $Rh^{3+}$ ($4d^6$) ion and the $Rh^{4+}$($4d^5$) ion, respectively. On the other hand, a trigonal distortion lowers the symmetry and lifts the degeneracy of the $t_2$-orbitals, so that the splitting $\Delta_p$ is reduced. Since $\Delta_p$ is much less than $\Delta_o$, a high spin state is more likely to arise, yielding S=2 for the $Rh^{3+}$ ion in the $RhO_6$ trigonal prism. Thus, the chain consisting of $Rh^{3+}$(o), $Rh^{3+}$(p), $Rh^{4+}$(o) and $Rh^{4+}$(o) is expected to correspond to a spin distribution of S=0, 2, ½ and ½. This scenario is consistent with the magnetic results presented below although other spin configurations cannot be ruled out. In addition, the Sr ions act to widely separate chains, resulting in an interchain distance of ~5.600 Å, nearly twice as long as the intrachain Rh-Rh distance, which precludes strong coupling between chains. Each chain is surrounded by six evenly spaced chains that form a triangular lattice in the ab-plane (see Fig.1a). All these structural features facilitate a coupled spin-chain system with a strong Ising character.

Shown in Fig.2 is the magnetic susceptibility $\chi$ as a function of temperature for (a) the c-axis ($\chi_c$) and the ab-plane ($\chi_{ab}$) at B=0.05 T and (b) $\chi_c$ at various fields. The most dominant feature is that the c-axis $\chi_c$ shows a sharp peak at $T_N$=23 K at B=0.05 T, indicating the presence of three-dimensional AFM order, i.e., the spin chains are primarily AFM coupled with each other. In contrast, the ab-plane $\chi_{ab}$ displays only a weak temperature dependence, as seen in Fig.2a. This large anisotropy underlines the dominant single-ion anisotropy associated with the CEF at the prismatic sites. It is noteworthy that $T_N$ is immediately followed by a shoulder or an anomaly at T*=21.5 K, which is only visible in low fields. T* accompanies the irreversibility upon in-field and zero-field cooling, which increases with decreasing T. Given the triangular lattice of the spin chains and the AFM coupling, such behavior may imply the existence of magnetic frustration of spins at T<T*.

A fit of high temperature data of $\chi_c$ for 80<T<350 K to a Curie-Weiss law yields an effective moment $\mu_{eff}$ of 7.3 $\mu_B$/f.u. and a positive Curie-Weiss temperature $\theta_{cw}$ of 28 K (see the inset). Deviations from the Curie-Weiss behavior occur below 45 K. The positive sign of $\theta_{cw}$ undoubtedly arises from the ferromagnetic character of the intrachain coupling. A weak AFM interchain interaction coupled with a strong FM intrachain interaction is thought to be the cause of magnetic frustration in the Co-based spin-chain systems [9, 12, 13]. It is noted that a large $\theta_{cw}/T_N$ (>10) is a characteristic for two- and three-dimensional frustrating lattices [27, 28]. Such a ratio of $\theta_{cw}/T_N$ (=1.16) for $Sr_5Rh_4O_{12}$ is nearly identical to that for the partial antiferromagnet $Ca_3Co_2O_6$ [2] but smaller than those of the three-dimensional frustrating lattices [28]. However, a large $\theta_{cw}/T_N$ is observed in the quasi-one-dimensional $Ca_5Ir_3O_{12}$ and $Ca_4IrO_6$, As can be seen below. Markedly, the phase transition at $T_N$ can be readily pushed to lower temperatures by increasing B and becomes ill-defined at around B=2 T (see Fig.2b).

Displayed in Fig.3 is the isothermal magnetization M(B) for (a) the c-axis ($M_c$) and the ab-plane ($M_{ab}$) at T=1.7 K and (b) the c-axis $M_c$ at various temperatures. The strong uniaxial anisotropy, a consequence of the Ising character of the spin-coupling, is illustrated as $M_{ab}$ and shows only weak linear field dependence and $M_c$ exhibits two step-like transitions. $M_c$ reveals a few features of the spin chains. First, for T<10 K, the saturation moment, $M_s$, reaches 5.30 $\mu_B$/f.u. at a critical field $B_c$=4.8 T. $M_s$ is close but slightly lower than the expected value of 6 $\mu_B$/f.u. for spin chains with spin configuration of S=0, 2, ½, and ½, assuming a Landé factor g=2. This discrepancy could be due to the inversion twinning at the Rh11 sites, which results in an average structure that superposes a trigonal prism and an octahedron. Thus, it is possible that the assumed high spin state at the Rh11 sites may be only partially realized, leading to a moment smaller than 4$\mu_B$ for S=2. Secondly, $M_c$ at T=1.7 K rises slightly but visibly at low fields (B<0.15 T) and then undergoes a sharp transition at $B^*$=2.4 T, reaching 1.73 $\mu_B$/f.u. or about $M_s$/3. After a rapid rise in an interval of 2.4 T, which is interestingly (but probably accidentally) equal to the value of B*, $M_c$ attains the value $M_s$ at $B_c$=4.8 T (see Fig.3a). While the rise in $M_c$ at low fields may be an indication of a slight lifting of degeneracy of the spin chains, the value of $M_s/3$ at B*=2.4 T is most likely a sign that the system enters a metamagnetic or ferrimagnetic state that contains FM chains with only 2/3 of them parallel to B and 1/3 anti-parallel to B, a situation somewhat similar to that of $Ca_3Co_2O_6$ [2, 9, 12]. Furthermore, the ferrimagnetic to FM transition at $B_c$ shows no hysteresis, suggesting that it is of second order. In contrast, hysteresis is pronounced below B* as shown in Fig.3a, which is indicative of a first order transition. This effect persists up to 23 K, but weakens as T rises. It reflects the character of a frozen spin state that prohibits full spin reversal when B ramps down to zero. No irreversibility would be expected if the magnetic order is purely AFM below B*. Clearly, this behavior emphasizes the existence of geometrical frustration for 0≤B<B*. With increasing T, B* decreases progressively whereas $B_c$ remains unchanged for T≤10 T and then increases slightly but broadens significantly for T>10 K, as seen in Fig. 3b. This suggests that the spin-flip process of the spin chains at B* is much more sensitive to the thermal energy than that at $B_c$, as expected for the quenching of the frustration by a field.

Fig. 4a illustrates the specific heat C as a function of temperature for 1.8≤T≤40 K. It exhibits an anomaly at $T_N$≈23 K, where ΔC~0.12 R (the gas constant R=8.31 J/mol K), confirming the existence of the long-range order at $T_N$. While the jump in C has the characteristic mean-field in shape, the broadened peak could be the consequence of the nearby second anomaly at T*=21.5 K immediately below $T_N$. It is remarkable that ΔC is rather small given the sharp phase transition seen in $\chi_c$. This small value of ΔC is then most likely a signature of the incomplete AFM ordering due to the geometrical frustration, consistent with our magnetic results. The plot of C/T vs $T^2$ shown in the upper inset displays a linear contribution to C, $\gamma T$, below 7 K, yielding $\gamma \sim 30$ mJ/mol K$^2$. Such sizable $\gamma$ in an insulator could arise from the excitations of a frustrated or disordered magnetic state at low T. Similar behavior is observed in disordered insulating magnets [26] and other frustrated systems. As T rises, C/T as a function of $T^2$ deviates from the linear dependence, possibly implying the emergence of different magnetic excitations (see both insets). These results further emphasize the presence of geometrical frustration due to the triangular lattice of spin chains at B=0. A finite $\gamma$ and a small $\Delta$C seem to be a common characteristic in spin-chain systems and seen as well in $Ca_3Co_2O_6$ (where $\gamma \sim 10$ mJ/mol-K$^2$ [13]), $Ca_5Ir_3O_{12}$ and $Ca_4IrO_6$ to be discussed below.

The B-T phase diagram in Fig.4b summarizes the magnetic properties of this frustrated Ising chain system. It is established that the system is antiferromagnetically ordered below 23 K with the intrachain and interchain couplings being FM and AFM, respectively. However, the 3D long-range AFM order at low fields is incomplete because of the triangular lattice formed by the spin chains that inevitably causes geometrical frustration. As B increases, the system enters a state with a partial FM order through a first order transition at B* and then the fully polarized FM state via a second order transition at $B_c$ (see Fig. 4b).

**B. Single crystal $Ca_5Ir_3O_{12}$**

$Ca_5Ir_3O_{12}$ is also a triangular lattice and shares a similar but somewhat less complex crystal structure compared to that of $Sr_5Rh_4O_{12}$. Refinements of the x-ray diffraction data reveal that $Ca_5Ir_3O_{12}$ adopts a hexagonal structure with the lattice parameters *a* and *c* being 9.4208 Å and 3.1941 Å, respectively. The space group is *P-62m* (187). The structure of the rod-shaped crystals features parallel chains of edge-sharing $IrO_6$ octahedra along the direction of the c-axis with no cross-linking between neighboring chains (the interchain Ir-Ir distance is 5.32 Å), as shown in Fig.5. Although there are one $Ir^{4+}$ ($4d^4$) ion and two $Ir^{5+}$ ($4d^3$) ions in each formula unit, the x-ray diffraction data show no crystallographical differences between the three Ir sites. These results are consistent with those published earlier [29-32]. The crystal structure of $Ca_5Ir_3O_{12}$ was initially identified as $Ca_2IrO_4$ [29]. But using the model in the reference 29, one of the Ca atoms refines in such a way that it essentially disappears. In fact, the reference 29 shows that the same Ca atom has very high vibrational parameter relative to the rest of the structure, suggesting either a wrong atom type assignment or less-than-full occupancy for the atom. The refinements of our x-ray diffraction data is in agreement with those in the reference 32 where polycrystalline samples were studied.

Fig. 6a shows the magnetic susceptibility as a function of temperature for the magnetic field B parallel ($\chi_{\parallel c\text{-axis}}$) and perpendicular ($\chi_{\perp c\text{-axis}}$) to the c-axis. It is clear that $\chi_{\parallel c\text{-axis}}$ exhibits a sharp magnetic phase transition at $T_M$=7.8 K and a strong hysteresis effect on zero-field cooling (ZFC) and field cooling (FC). The behavior mimics that of a ferromagnet but the isothermal magnetization and other magnetic measurements, as shown below, suggest that the ferromagnetic-like behavior is only a reflection of a canted antiferromagnetic state. The irreversibility gets weaker and eventually vanishes as B increases. Remarkably, $\chi_{\perp c\text{-axis}}$ is larger than $\chi_{\parallel c\text{-axis}}$ but shows only a weak anomaly at $T_M$. Apparently, the single ion anisotropy due to the crystal field is not sufficient enough to generate Ising spins such as those seen in $Sr_5Rh_4O_{12}$. Furthermore, as B increases, the transition is broadened but not suppressed to lower temperatures as shown in Fig.6b. This behavior is also different from that observed in $Sr_5Rh_4O_{12}$ (Fig.2b) and unusual in that a Neel temperature is expected to decrease with increasing B and a field of 6 T should be sufficient to suppress a Neel temperature such as 7.8 K to nearly zero for a conventional antiferromagnet. Such broadening of $T_M$ with increasing B resembles the ferromagnetic behavior, but both the Curie-Weiss analysis of high- temperature $\chi$ and the isothermal magnetization confirm the presence of the AFM state.

Shown in Fig.7a is the inverse magnetic susceptibility at B=0.2 T $\chi_{\parallel c}^{-1}$ and $\chi_{\perp c}^{-1}$ as a function of temperature. The fitting data to the Curie-Weiss law for 40<T<350 K yields a Curie-Weiss temperature $\theta_{cw}$ to be -280 K and -102 K for $\chi_{\parallel c}^{-1}$ and $\chi_{\perp c}^{-1}$, respectively. While the negative sign of $\theta_{cw}$ clearly indicates the AFM coupling, the large magnitudes of $\theta_{cw}$ make the ratio of $\theta_{cw}/T_M$ (>35 and 13 for $\chi_{\parallel c}^{-1}$ and $\chi_{\perp c}^{-1}$, respectively) unusually high for any conventional antiferromagnets. Since $\theta_{cw}$ is a measure of the exchange coupling, a phase transition to a long-range antiferromagnetic order is expected to occur at T~$\theta_{cw}$ when the exchange coupling is strong enough to overcome thermal fluctuations, according to the mean field theory [33]. (For regular antiferromagnets, the experimentally determined $\theta_{cw}$ could be a few times larger than a Neel temperature. The discrepancy is largely due to the assumption that the molecular field on one sublattice depends only on the magnetization of the other sublattice). Large values of the ratio of $\theta_{cw}/T_M$ such as these presented are observed in pyrochlore and kagome systems, as pointed out above [27, 28]. The ratio $\theta_{cw}/T_M$ is often seen as a measure of the magnetic frustration in two- and three-dimensional frustrating lattices, and a large value of $\theta_{cw}/T_M$ thus corresponds to a large depression of the phase transition temperature. It is conceivable that despite the large $\theta_{cw}$ or strong antiferromagnetic interaction, the phase transition occurs only at $T_M$=7.8 K due to the geometric frustration that prevents the long-range magnetic ordering from occurring until at a temperature well below $\theta_{cw}$. The linearity in $\chi_{\|c}^{-1}$ and $\chi_{\perp c}^{-1}$ at $T<|\theta_{cw}|$ implies the presence of the nearest-neighbor interaction frequently seen in a situation of a frustrated system [28]. According to mean field theory, the linearity of the inverse magnetic susceptibility is expected only at $T>>|\theta_{cw}|$ for a conventional magnet. It is noted that the low dimensionality does not favor a long-range order in general and may also cause relatively large values of $\theta_{cw}/T_M$ but normally smaller than 10. The behavior displayed in Fig.7a is similar to that of geometrically frustrated magnets such as $CsMnFeF_6$ [28, 34] whereas $\chi_{\|c}^{-1}$ is rather linear in temperatures down to 60 K, well below $\theta_{cw}$(=280 K). Isothermal magnetization at T=1.7 K in Fig.7b shows a linear field dependence of M for both orientations, consistent with the behavior for an AFM state.

The heat capacity C(T) as a function of temperature as shown in Fig.8a displays a peak corresponding to $T_M$=7.8 K, confirming the 2$^{nd}$ order phase transition. The peak is relatively broad and gives $\Delta C \sim 0.23$ R, too small for complete spin-ordering. Moreover, the plot of C/T vs $T^2$ as shown in the inset also exhibits a linear contribution $\gamma T$ to C(T) below 5 K, yielding $\gamma \sim 76$ mJ/mol $K^2$. This value is even large than that for $Sr_5Rh_4O_{12}$, signaling once again the excitations of a frustrated or disordered magnetic state at low temperatures in an insulating ground state. Subtracting a baseline obtained by fitting the data for 20<T<40 K to a polynomial gives the magnetic contribution $\Delta C_M$, and integrating $\int \Delta C_M/T dT$ yields the corresponding entropy removal $\Delta S$ as shown in Fig.8b (right scale). $\Delta S$ is approximately 4.30 J/mole K, smaller but reasonably close to Rln2 or 5.76 J/mole K expected for complete ordering of S=1/2 spins. Since $Ca_5Ir_3O_{12}$ involves both a low-spin state S=1/2 ($Ir^{4+}$) and S=3/2 ($Ir^{5+}$), the value of $\Delta S$ suggests that only S=1/2 spins of the $Ir^{4+}$ ions participate in the magnetic ordering whereas S=3/2 spins of the $Ir^{5+}$ ions remain disordered at low temperatures. In addition, C/T near $T_M$ changes only slightly upon the application of a relatively strong field up to 9 T, implying a small reduction of entropy, consistent with the behavior seen in χ (Fig. 6). The insensitivity of $T_M$ to the magnetic field might suggest an occurrence of competing AFM and FM interactions motivated by the magnetic field.

It is interesting that electrical resistivity perpendicular to the c-axis $\rho_{\perp c}$ appears to show a crossover from a metallic state to an insulating state near 300 K as temperature decreases whereas resistivity parallel to the c-axis $\rho_{\parallel c}$ displays variable range behavior throughout the entire temperature range of 50<T<700 K, as can been seen in Fig. 9. ρ rapidly increases by more than five orders of magnitude from $10^{-2}$ Ω cm at 700 K to $10^3$ Ω cm at 65 K. log $\rho_{\parallel c}$ at high temperatures approximately obeys temperature dependence of $T^{-1/2}$, a power law often expected for quasi-one-dimensional systems. But the $T^{-1/2}$-dependence is not so well-defined for log $\rho_{\perp c}$ (see the inset). It is also noted that there is a slope change near 100 K (see the inset) however no magnetic anomaly is discerned in χ.

**C. Single crystal $Ca_4IrO_6$**

$Ca_4IrO_6$ crystallizes in a rhombohedral structure with *a*=9.3030 Å, *c*=11.0864 or a $K_2CdCl_6$ type rhombohedral structure with space group of R-3c (167). The crystal structure consists of one-dimensional chains of alternating $IrO_6$ octahedra and $CaO_6$ trigonal prisms running parallel to the c-axis. It is otherwise identical to that of the Ising chain $Ca_3Co_2O_6$ [2]. The crucial difference between $Ca_3Co_2O_6$ and $Ca_4IrO_6$ is that the magnetic $CoO_6$ trigonal prisms running parallel to the c-axis in the former are replaced by the non-magnetic $CaO_6$ trigonal prisms in the latter. This crystal structure was first reported in an early study [31] and confirmed by a recent study [35] and this work. These crystals are rod-shaped and involve only $Ir^{4+}$ ions with a low spin state S=1/2.

Fig.10a displays the magnetic susceptibility $\chi_{\|c}^{1}$ and $\chi_{\perp c}$ for B=0.5 T as a function of temperature for both the c-axis and ab-plane. The broad peaks seen near 12 K indicate an AFM phase transition. The temperature dependence of $\chi$ is similar for both orientations, in contrast to that of $Sr_5Rh_4O_{12}$ and $Ca_5Ir_3O_{12}$. Fits of high-temperature (200<T<350 K) $\chi_{\|c}$ and $\chi_{\perp c}$ to the Curie-Weiss law yield $\theta_{cw}$ to be 60 K and 32 K, respectively. The ratio $\theta_{cw}/T_M$ is smaller than that for $Ca_5Ir_3O_{12}$ but $\chi_{\|c}^{-1}$ is rather linear in temperature for T<$\theta_{cw}$ as shown in Fig.10b. As discussed above, this reflects the interaction between the nearest neighbors expected for frustrated systems. The temperature dependence of $\chi_{\|c}^{-1}$ and $\chi_{\perp c}^{-1}$ resembles that of $CsMnFeF_6$ [28, 34]. Isothermal magnetization M at T=1.7 K in Fig.11 shows a linear field dependence of M for both orientations, verifying the AFM state. It is noted that M for $Ca_4IrO_6$ shows stronger field dependence than that for $Ca_5Ir_3O_{12}$. Although closely related to $Ca_3Co_2O_6$ structurally, $Ca_4IrO_6$ does not show the step transitions in M that characterize the former [2, 9]. It is interesting that the step-like magnetization only occurs in frustrating lattices such as $Ca_3Co_2O_6$ and $Sr_5Rh_4O_{12}$ where the spin chains of ($CoO_6$ or $RhO_6$) trigonal prisms running parallel to the c-axis accompany spin chains of ($CoO_6$ or $RhO_6$) octahedra. The trigonal prisms, as discussed above, reduce the symmetry, thus the splitting $\Delta_p$ ($<\Delta_o$), therefore a high spin state is more likely to arise. The absence of the step-like magnetization in both $Ca_4IrO_6$ and $Ca_5Ir_3O_{12}$ seems to be correlated with the absence of the spin chains of trigonal prisms despite the crystal structures highly similar to those of $Ca_3Co_2O_6$ and $Sr_5Rh_4O_{12}$.

Shown in Fig.12a is the heat capacity C(T) as a function of temperature. C(T) shows a peak at $T_M$=12 K, indicating the 2nd order phase transition. But such a magnetic phase transition induces only a small $\Delta C$~0.11 R, once again implying incomplete spin-ordering. In addition, the plot of C/T vs $T^2$ (see the inset) also shows a finite linear contribution $\gamma T$ to C(T) below 9 K with $\gamma$~24 mJ/mol $K^2$, an indication of the excitations of a frustrated magnetic state at low T. The application of the magnetic field up to 5 T exerts no significant impact on the phase transition as shown in Fig.12b where C/T vs T is plotted for B=0 and 5 T. This behavior is similar to that of $Ca_5Ir_3O_{12}$, signaling competing AFM and FM interactions.

## IV. Conclusions

These quasi-one-dimensional materials are distinguished from each other through interesting differences in magnetic behavior. $Sr_5Rh_4O_{12}$ is defined by a strong Ising characteristic and the two step-like transitions in the c-axis isothermal magnetization that lead to a ferrimagnetic state with 1/3 of M at $B^*$=2.4 T and a fully saturated FM state at $B_c$=4.8 T. It appears that the step-like magnetization is critically associated with the spin chains of trigonal prisms running parallel to the c-axis, which are unique to $Sr_5Rh_4O_{12}$. Both $Ca_5Ir_3O_{12}$ and $Ca_4IrO_6$ are also characterized by the AFM order below 7.8 K and 12 K, respectively, and exhibit a large $\theta_{cw}/T_M$, implying a significantly suppressed ordered state or geometric frustration. Furthermore, $Ca_5Ir_3O_{12}$ illustrates that only S=1/2 spins of the $Ir^{4+}$ ions are involved in the magnetic ordering whereas S=3/2 spins of the $Ir^{5+}$ ions freezes in a disordered state. In spite of the varied magnetic behavior, these materials share common characteristics central to geometric frustration. They feature crystal structures that favor the formation of spin chains and a triangular lattice. As a result, they are antiferromagnetically ordered with incomplete spin ordering. The partial AFM state is clearly evidenced in the small ΔC and/or the entropy removal associated with the phase transitions and the finite low-temperature specific heat linearly proportional to temperature in spite of the insulating ground state. These intriguing phenomena raise interesting questions and we hope this work will stimulate more investigations on these one-dimensional 4d- and 5d-based materials that offer a wide window into low-dimensional magnetism involving geometrically frustrated states.

**Acknowledgement:** This work was supported by NSF grants DMR-0240813, DMR-0552267 and DOE grant DE-FG02-98ER45707.

**Captions:**

**Fig.1.** (a) The projection of the crystal structure on the ab-plane, and (b) chain arrays along the c-axis of $Sr_5Rh_4O_{12}$. The large solid circles are Rh ions; the dark small circles oxygen ions and the gray small circles Sr ions. The following is the Rh-Rh bond distance for the three chains: 2.747 Å (Rh4-Rh1), 2.744 Å (Rh1-Rh2), 2.578 Å (Rh2-Rh3), 2.586 Å (Rh3-Rh4); 2.601 Å (Rh8-Rh5), 2.543 Å (Rh5-Rh6), 2.753 Å (Rh6-Rh7), 2.758 Å (Rh7-Rh8); 2.604 Å (Rh12-Rh9), 2.580 Å (Rh9-Rh10), 2.732 Å (Rh10-Rh11), 2.739 Å (Rh11-Rh12).

**Fig.2.** The magnetic susceptibility $\chi$ of $Sr_5Rh_4O_{12}$ as a function of temperature for (a) the c-axis ($\chi_c$) and the ab-plane ($\chi_{ab}$) at B=0.05 T, and (b) $\chi_c$ vs. T at various fields. Inset: $\Delta\chi^{-1}$ vs T ($\Delta\chi$ is defined as $\chi-\chi_o$ where $\chi_o$ (~0.016 emu/mole) is the temperature-independent contribution to $\chi$.

**Fig.3.** The isothermal magnetization M(B) of $Sr_5Rh_4O_{12}$ for (a) the c-axis ($M_c$) and the ab-plane ($M_{ab}$) at T=1.7 K and (b) the c-axis $M_c$ at various temperatures. The dots sketch the hexagonal lattice of spin chains with solid dots and empty dots corresponding to spins parallel and antiparallel to B, respectively. Note that the value of $M_c$ at B* is 1/3 of $M_s$ at $B_c$ for T=1.7 K.

**Fig.4.** (a) The specific heat C of $Sr_5Rh_4O_{12}$ as a function of temperature for $1.8 \leq T \leq 40$ K. Insets: C/T vs. $T^2$ for 0<T<13 (upper) and 0<T<40 (lower); (b) The B-T phase diagram generated based on the data in Fig.3.

**Fig.5.** (a) The projection of the crystal structure on the ab-plane, and (b) chain arrays along the c-axis of $Ca_5Ir_3O_{12}$.

**Fig.6.** The magnetic susceptibility $\chi$ of $Ca_5Ir_3O_{12}$ as a function of temperature for (a) parallel ($\chi_{||c-axis}$) and perpendicular ($\chi_{\perp c-axis}$) to the c-axis at B=0.05 T, and (b) $\chi_{||c-axis}$ and $\chi_{\perp c-axis}$ vs. T at B=1 and 6 T.

**Fig.7.** (a) The inverse magnetic susceptibility $\chi_{||c}^{-1}$ and $\chi_{\perp c}^{-1}$ of $Ca_5Ir_3O_{12}$ as a function of temperature; (b) The isothermal magnetization for $M_{||c}$ and $M_{\perp c}$ at T=1.7 K.

**Fig.8.** (a) The specific heat C of $Ca_5Ir_3O_{12}$ as a function of temperature for $1.8 \leq T \leq 40$ K. Inset: C/T vs. $T^2$ for 0<T<5; (b) C/T vs T for B=0, 5, and 9 T (left scale) and the entropy removal $\Delta S$ vs T (right scale).

**Fig.9.** (a) The magnetic susceptibility $\chi$ of $Ca_4IrO_6$ as a function of temperature for $\chi_{||c-axis}$ and $\chi_{\perp c-axis}$ at B=0.05 T, and (b) the inverse magnetic susceptibility $\chi_{||c}^{-1}$ and $\chi_{\perp c}^{-1}$ as a function of temperature. The dashed lines are a guide to the eye.

**Fig.10.** Electrical resistivity parallel and perpendicular to the c-axis, $\rho_{\|c}$ and $\rho_{\perp c}$, of $Ca_5Ir_3O_{12}$ as a function of temperature for 50<T<700 K. Inset: log r vs. for $T^{-1/2}$ for $\rho_{\|c}$ and $\rho_{\perp c}$.

**Fig.11.** The isothermal magnetization of $Ca_4IrO_4$ for $M_{\|c}$ and $M_{\perp c}$ at T=1.7 K.

**Fig.12.** (a) The specific heat C of $Ca_4IrO_6$ as a function of temperature for 1.8≤T≤ 40 K. Inset: C/T vs. $T^2$ for 0<T<9; (b) C/T vs T for B=0 and 5 T.

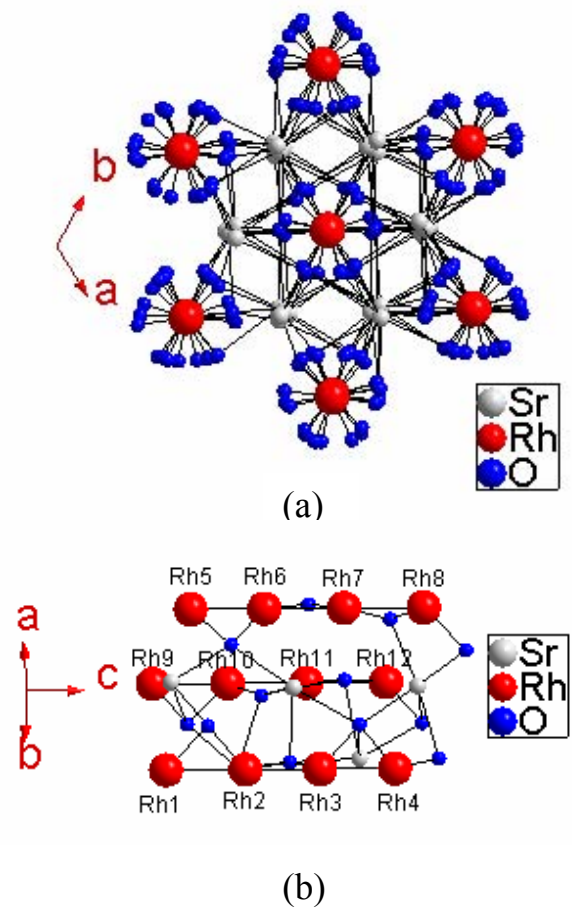

Fig.1

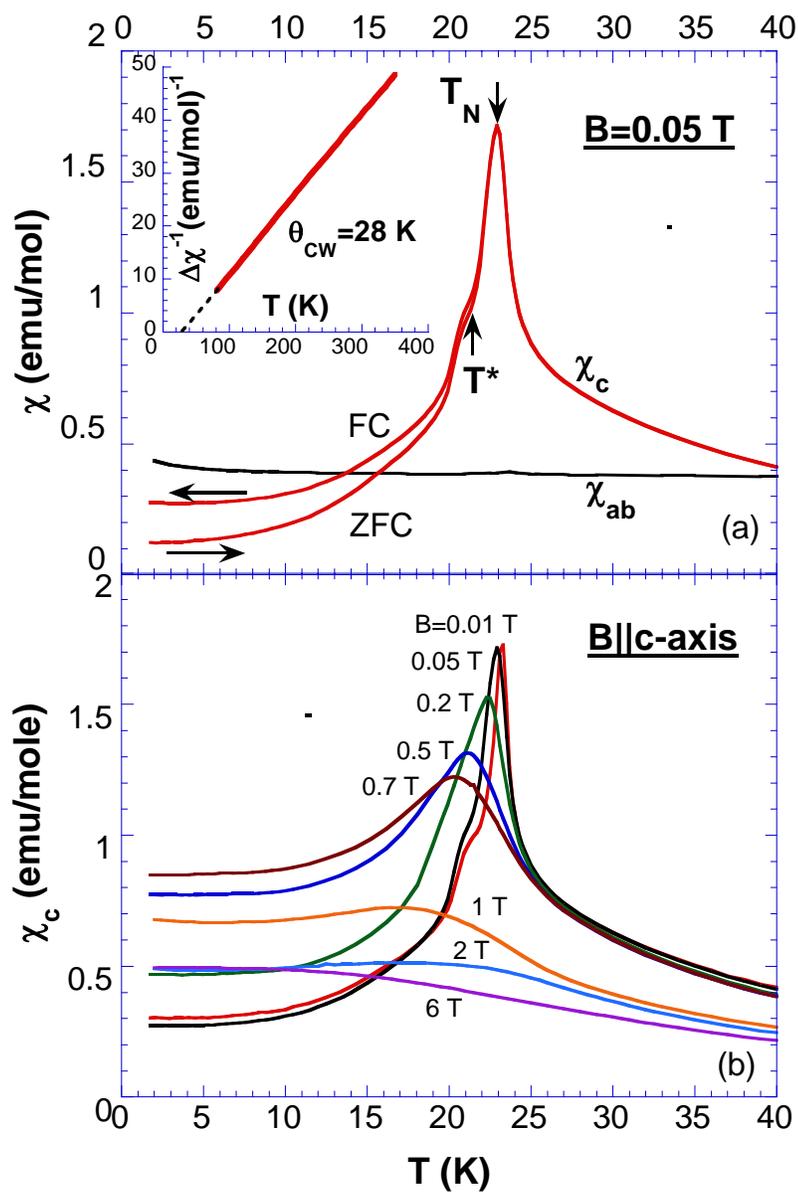

Fig.2

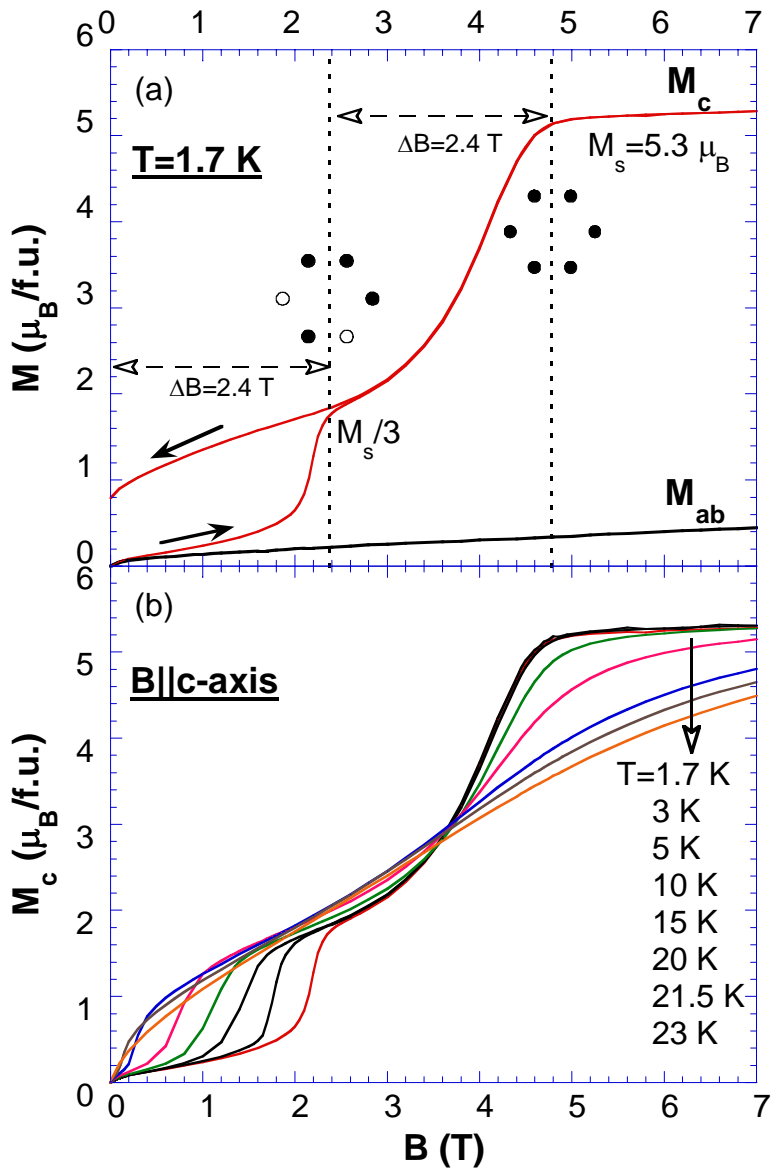

Fig.3

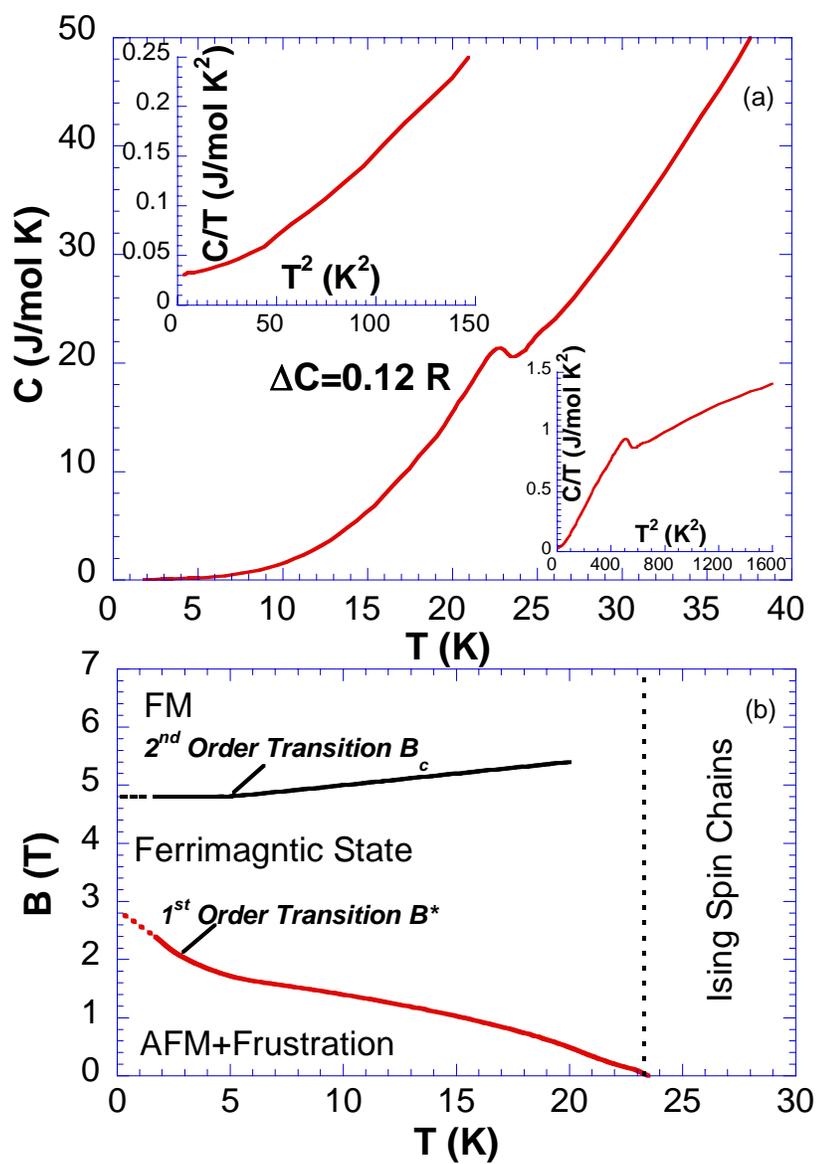

Fig. 4

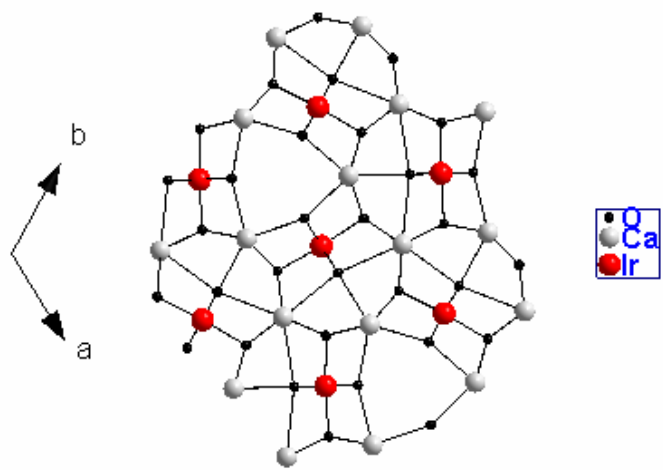

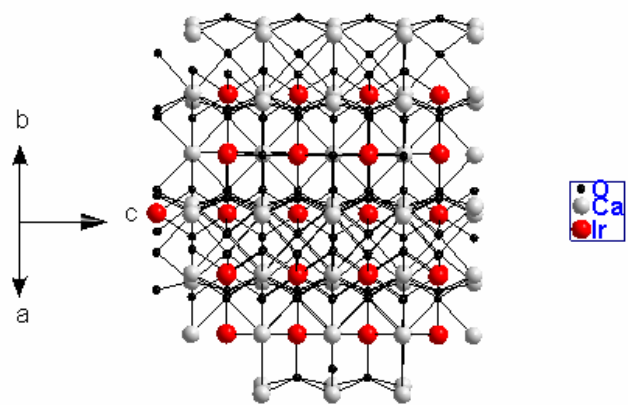

Fig.5

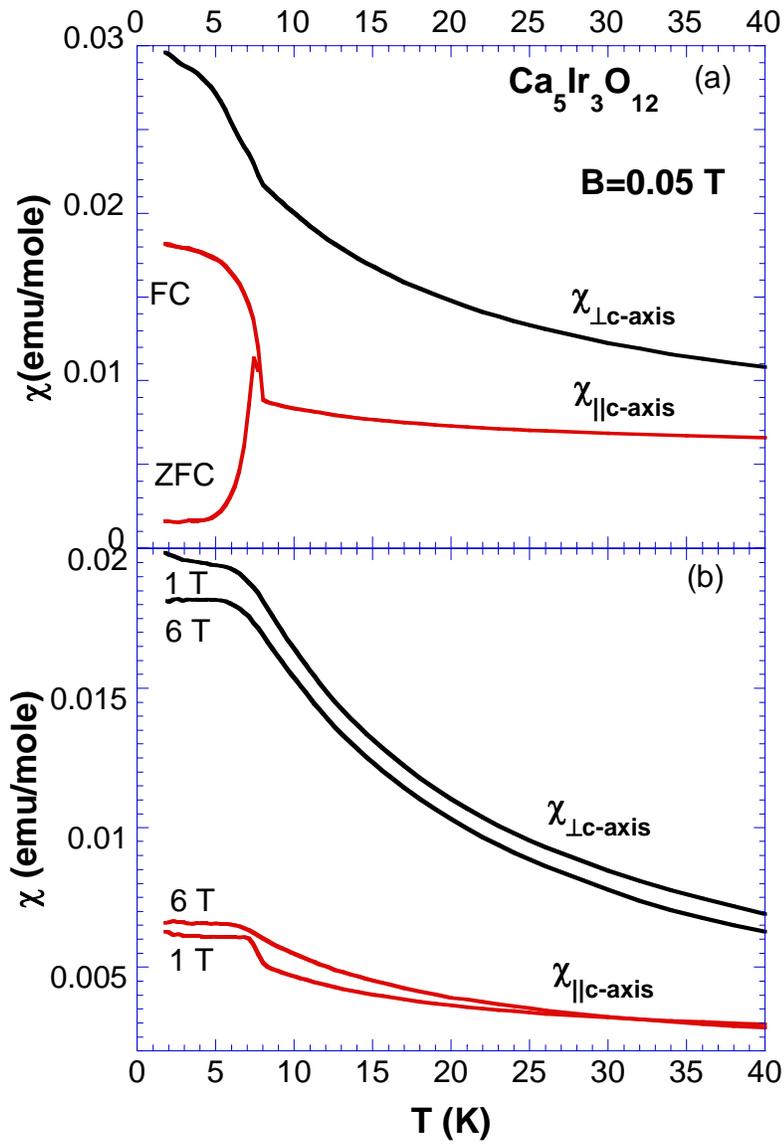

Fig. 6

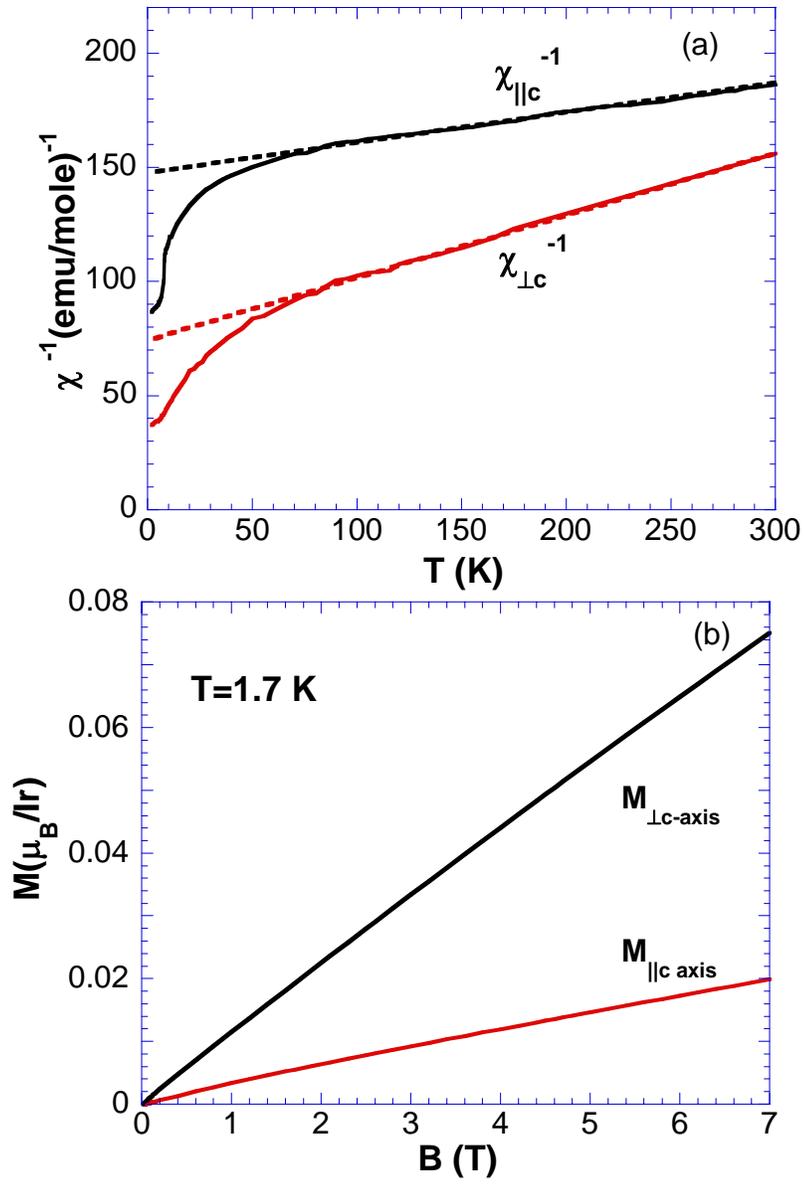

Fig. 7

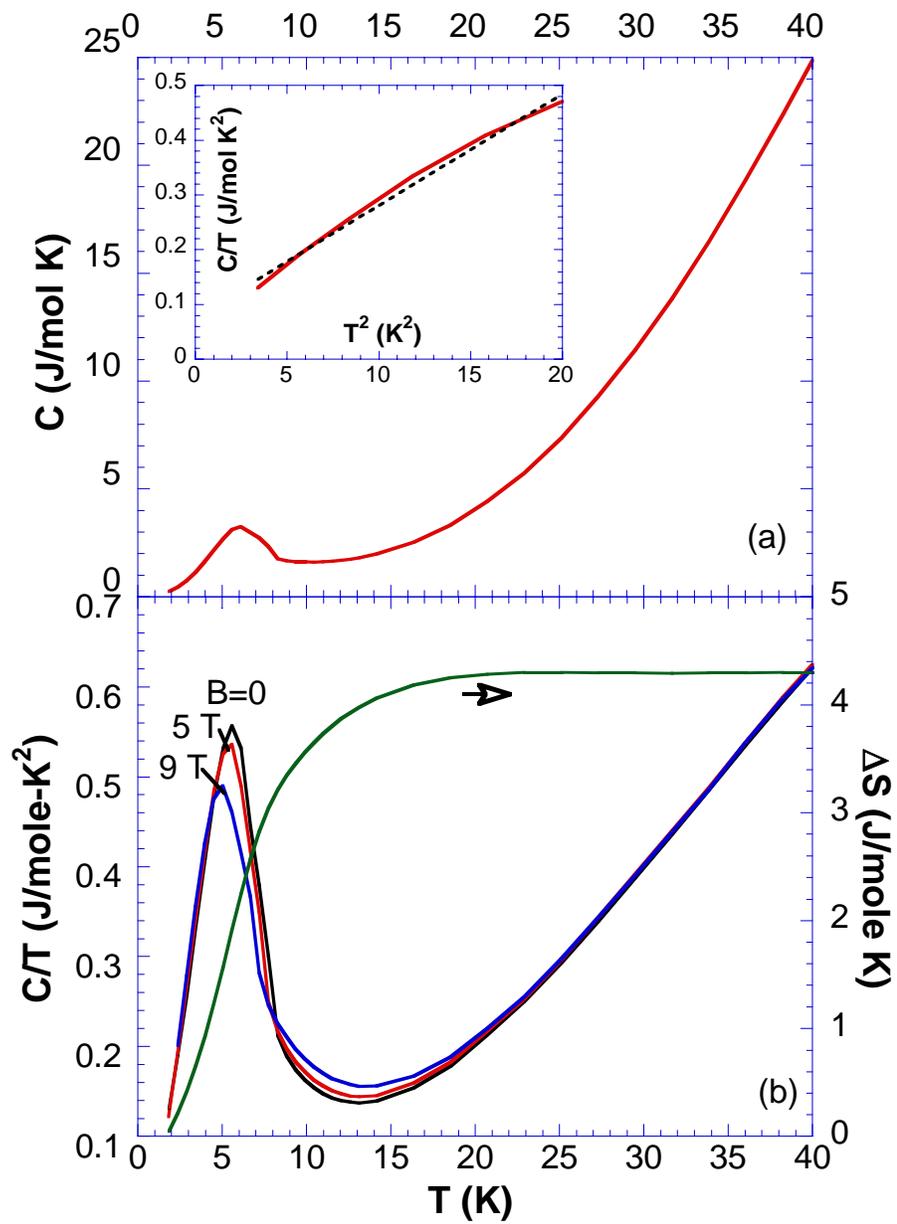

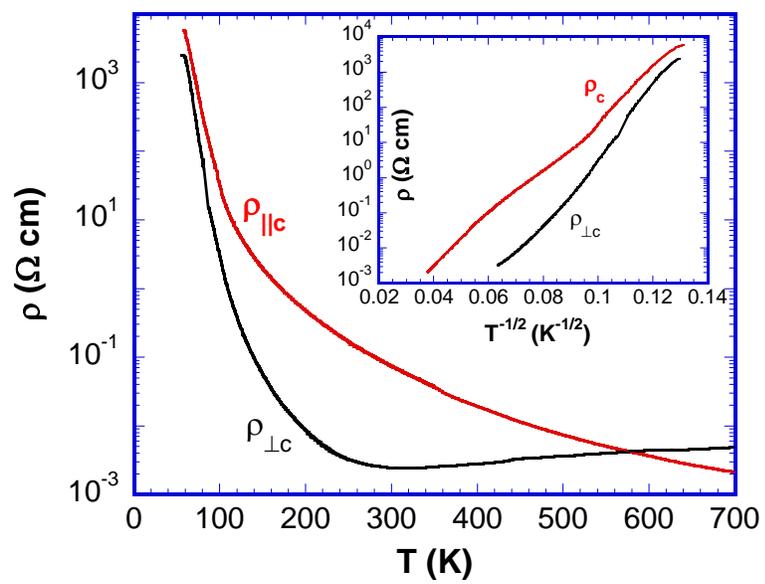

Fig. 9

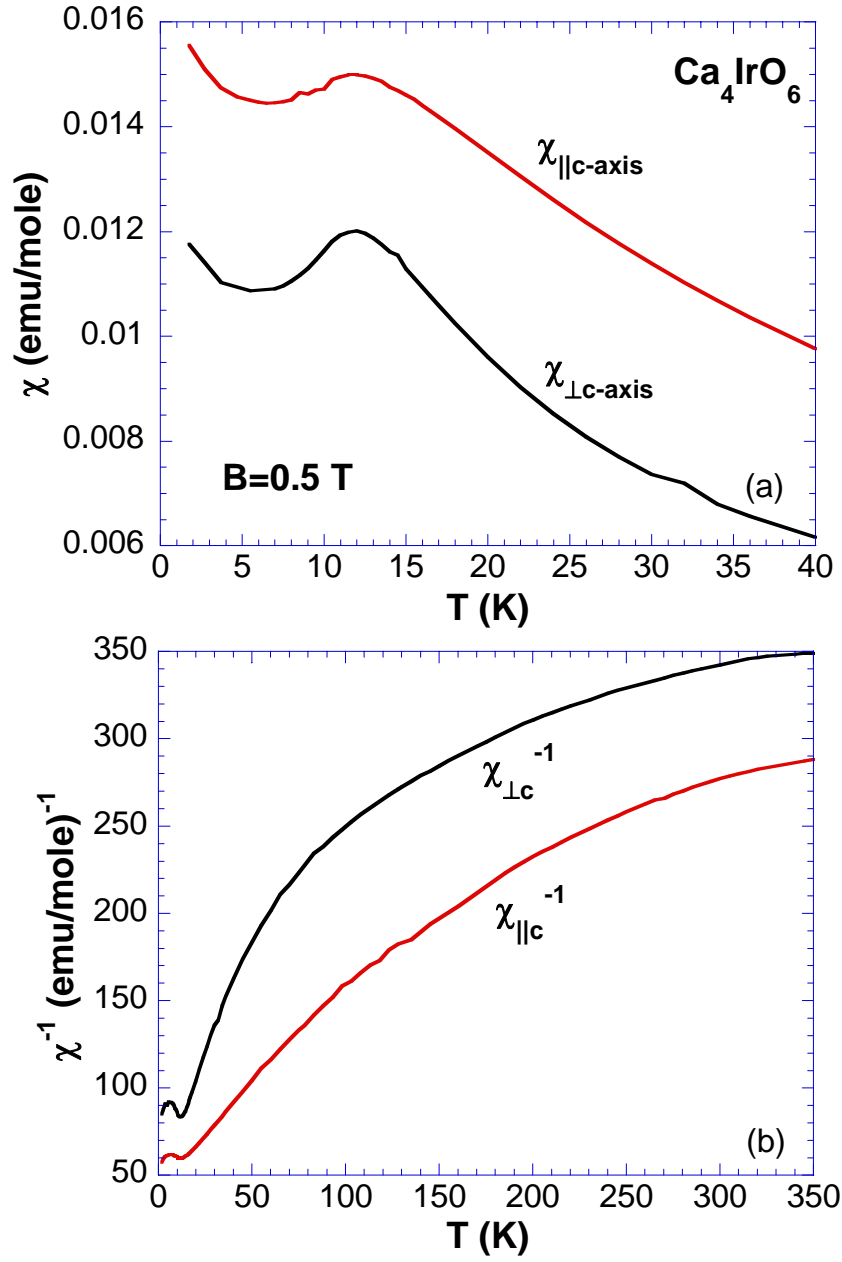

Fig. 10

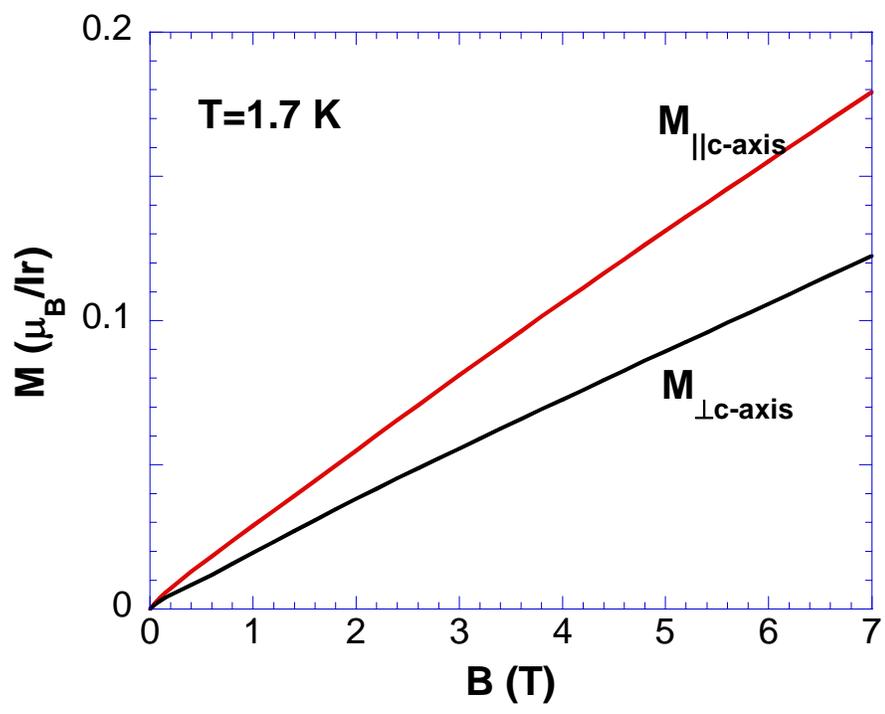

Fig. 11

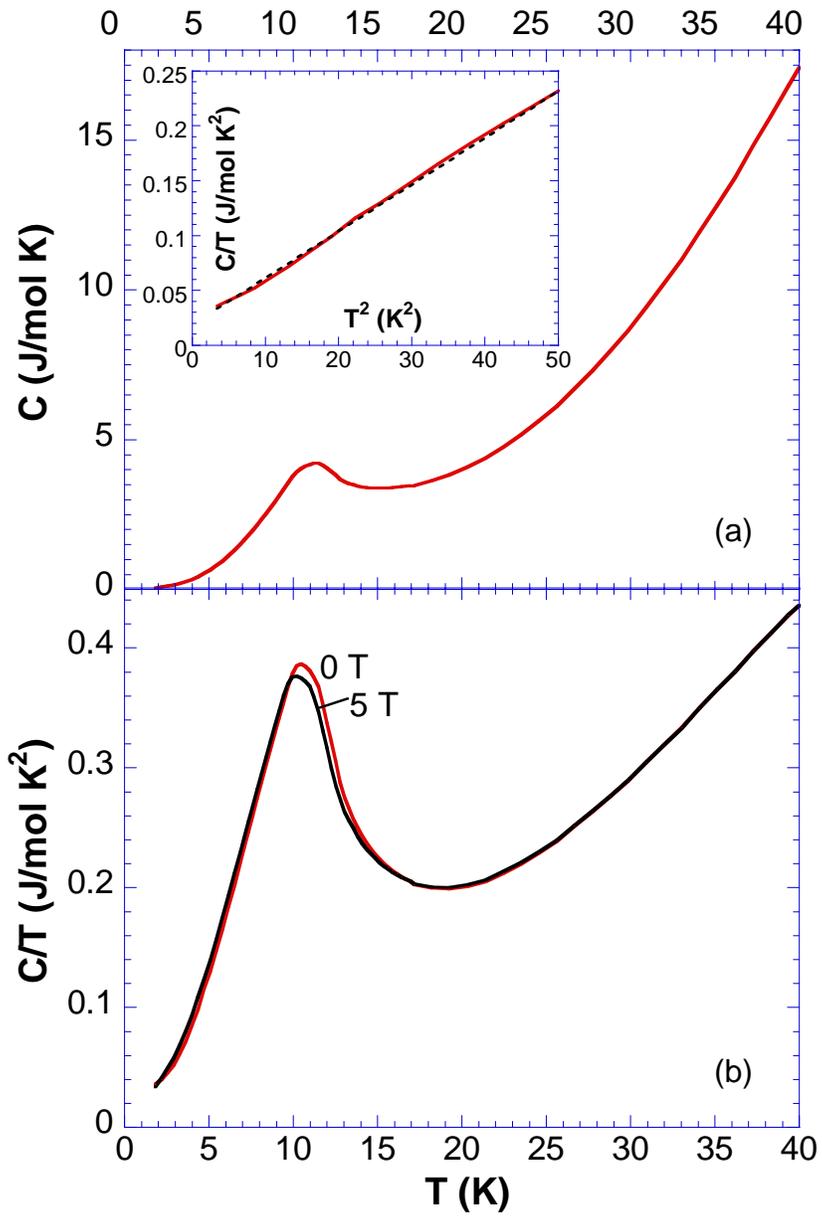

Fig. 12